\documentclass[twocolumn,twocolappendix]{aastex631}
\usepackage{amsmath,textcomp,gensymb}
\usepackage{xcolor}
\shortauthors{Behmard et al.}
\graphicspath{{./}{figures/}}

\begin{document}

\title{A Link Between Rocky Planet Density and Host Star Chemistry}

\author[0000-0003-0012-9093]{Aida Behmard}
\altaffiliation{Flatiron Research Fellow}
\affiliation{Center for Computational Astrophysics, Flatiron Institute, 162 Fifth Ave, New York, NY 10010, USA}
\affiliation{American Museum of Natural History, 200 Central Park West, Manhattan, NY 10024, USA}

\author[0000-0002-4480-310X]{Casey L. Brinkman}
\affiliation{Department of Physics and McGill Space Institute, McGill University, 3600 University Street, Montreal, QC H3A 2T8, Canada}

\author[0000-0002-0842-863X]{Soichiro Hattori}
\affiliation{Department of Astronomy, Columbia University, 538 West 120th Street, Pupin Hall, New York, NY 10027, USA}
\affiliation{American Museum of Natural History, 200 Central Park West, Manhattan, NY 10024, USA}

\author[0000-0003-3856-3143]{Ryan A. Rubenzahl}
\affiliation{Center for Computational Astrophysics, Flatiron Institute, 162 Fifth Ave, New York, NY 10010, USA}


\author[0000-0001-9907-7742]{Megan Bedell}
\affiliation{Center for Computational Astrophysics, Flatiron Institute, 162 Fifth Ave, New York, NY 10010, USA}


\begin{abstract}
Planets and their host stars form from the same cloud of gas and dust, so we assume that their chemical compositions are linked. However, a clear correlation between rocky planet interior properties and host star chemistry remains elusive for planets around FGK dwarfs, and non-existent for planets around M dwarfs because cool stars frequently lack detailed chemical information. Here, we investigate the relationship between small ($R_{P}$ $\leq$ 1.8 $R_{\oplus}$) planet densities and host star elemental abundances. We use the Sloan Digital Sky Survey-V/Milky Way Mapper and an accompanying data-driven framework to obtain abundances for FGK and M dwarf hosts of 22 rocky planets. We find that planet densities exhibit a strong, inverse relationship to [Mg/Fe] abundances of FGK hosts ($p$ = 0.001). This correlation becomes more significant with the addition of M dwarf hosts ($p$ = 0.0005). If we assume that rocky planets have terrestrial-like compositions, this suggests that low [Mg/Fe] environments form planets with larger Fe-rich cores and thus higher densities. The thick disk planets in our sample help anchor this trend, illustrating the importance of sampling exoplanet properties across a range of host star populations. This finding highlights the connection between Galactic chemical evolution and rocky planet formation, and indicates that Earth-like planet compositions may vary significantly across different regions of the Galaxy. 


\end{abstract}



\section{Introduction} \label{sec:intro}
Planets and their host stars form from the same nebular material, so we expect that their abundances of solid, rock-forming elements will be roughly commensurate. This is true for most rocky planets in the solar system. To first order, the Earth is a devolatilized piece of the solar nebula, with refractory element abundances indistinguishable from solar abundances \citep{wang2019b}. Like Earth, Mars has an Fe/Mg ratio that agrees to within $\sim$10$-$15\% of the Sun's abundances (e.g., \citealt{wang2019b,lodders2003,mcdonough2003}), and while Fe/Mg is unconstrained for Venus, it also has an Earth-like bulk composition \citep{zharkov1983}. Additionally, CI chondrites---rocky bodies that have undergone minimal chemical processing since the formation of the solar system---have detailed refractory abundance patterns that agree with those of the Sun to within 10\% \citep{lodders2003}, and Fe/Mg and Si/Mg that agree to within 2\% and 4\%, respectively \citep{putirka2019}. 

However, other solar system rocky bodies exhibit large abundance differences compared to the Sun. Mercury has an Fe-to-silicates concentration that is $\sim$200$-$400 times greater than solar composition (e.g., \citealt{morgan1980}). Conversely, the moon is Fe-depleted, with only $\sim$3\% of its mass contributing to an iron core (as opposed to $\sim$30\% for the Earth) \citep{szurgot2015}. Their deviations from solar composition may have resulted from chemical processing and/or different formation mechanisms. For example, Mercury's high iron abundance may be the result of evaporative silicate loss following a giant impact (e.g., \citealt{benz2008,chau2018}), or formation from preferentially Fe-rich planetesimals (e.g., \citealt{weidenschilling1978,anders2022}). The low Fe content of the moon may also result from a giant impact, where the moon formed from Fe-poor debris after a catastrophic collision with Earth \citep{canup2001}.

If rocky exoplanets have diverse formation pathways and/or often undergo significant chemical processing, we expect that their compositions will not reflect those of their host stars. Whether or not this is true has implications for rocky planet properties on large scales, because stellar chemistry is fundamentally set by Galactic chemical evolution. At present, we have discovered that rocky ($R_{P}$ $\leq$ 1.8 $R_{\oplus}$) exoplanet bulk densities span a wide range, beyond that of solar system rocky bodies (e.g., \citealt{morton2016,dai2019}). These densities are consistent with chemical compositions that could be iron-dominated (e.g., \citealt{bonomo2019}), composed entirely of silicates, or characterized by substantial water enrichment (e.g., \citealt{angelo2017,kite2021}). It is unclear if such possible planet compositions correlate with host star chemistry. \citet{adibekyan2021} report a star-planet compositional link, in this case between rocky planet iron-mass fraction-like quantities and host star elemental abundances. However, subsequent studies with larger, more homogeneous samples find that this correlation is less clear and highly dependent on fitting method (e.g., \citealt{brinkman2024}). 

A major challenge of these investigations is building a rocky planet sample with high-quality density measurements. Rocky planets have small sizes and masses that are difficult to precisely measure. Previous studies report that these uncertainties, particularly in mass, often confound efforts to measure planet densities and constrain interior compositions (e.g., \citealt{plotnykov2020,schulze2021}). However, progress is being made thanks to recent transit and radial velocity (RV) surveys that are larger and more precise than their predecessors, which are enabling us to construct rocky planet samples with mass and radius measurement uncertainties of $<$30\% (e.g., \citealt{gillon2017,dai2019,agol2021,thygesen2023,bonomo2023,polanski2024,brinkman2025}).



Constraining host star compositions is another difficult task. Stellar abundances are traditionally measured with stellar atmosphere models and spectral synthesis pipelines. These methods are well-suited to solar-like stars, and are often used to derive elemental abundances for FGK dwarf planet hosts (e.g., \citealt{adibekyan2021,brinkman2024,adibekyan2024b}). However, measuring abundances for M dwarf hosts is more challenging. Stellar atmosphere models lack the necessary spectral line lists/opacity information for reproducing complex molecular features that heavily populate cool star spectra (e.g., \citealt{mann2013b}). Additionally, many of the strongest, unblended M dwarf spectral lines reside in near-infrared regions, making them inaccessible via optical spectrographs often used in exoplanet surveys. This is unfortunate because M dwarfs are ideal for rocky planet discoveries---they are the most common stars in the Galaxy, and their small sizes and masses result in strong transit and RV signals, even for small, Earth-sized worlds (e.g., \citealt{henry2006,gillon2017,reyle2021}). Thus, developing robust methods for characterizing M dwarf chemistry is necessary for understanding the rocky planet population.

In this study, we examine potential correlations between rocky planet properties and host star compositions for both FGK and M dwarf hosts. This is made possible by a catalog of M dwarf abundances inferred from a data-driven approach \citep{behmard2025}. We outline our rocky planet sample selection in Section \ref{sec:sample}, including our procedure for homogeneously re-deriving planet masses and radii, and how we source our host star abundance information from the Sloan Digital Sky Survey-V/Milky Way Mapper (SDSS-V/MWM). In Section \ref{sec:correlation}, we explore connections between rocky planet properties and host star abundances for elements common in bulk Earth composition, and uncover a significant correlation between planet bulk densities and host star [Mg/Fe]. We discuss this finding in Section \ref{sec:discussion}, and the connections it implies between the rocky planet population and Galactic chemical evolution. 


\begin{figure*}[t]
    \centering
    \includegraphics[width=0.988\textwidth]{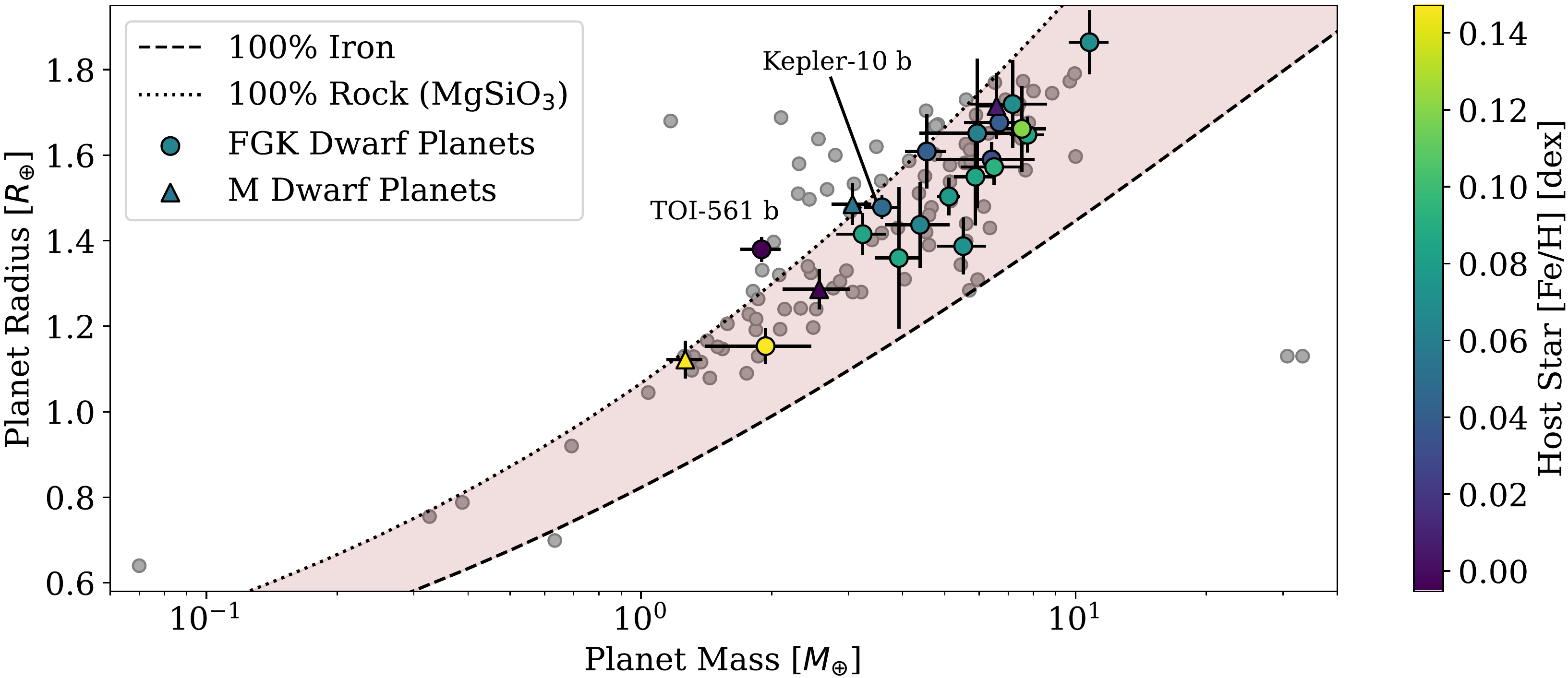} 
\caption{All small ($R_{P}$ $\leq$ 1.8 $R_{\oplus}$), rocky planets from the Exoplanet Archive \citep{christiansen2025} with $<$30\% fractional radius and mass uncertainties (gray points). Planets with masses and radii consistent with a rocky composition (pure iron to pure MgSiO$_{3}$ rock) are contained within the brown shaded region \citep{zeng2019}. Our rocky planet sample with host star abundances from SDSS-V are shown in color, which corresponds to host star [Fe/H]. Planets hosted by FGK dwarfs are marked as circles, while planets hosted by M dwarfs are marked as triangles. The one planet not contained within the \citet{zeng2019} region is TOI-561 b, the only high-confidence thick disk planet in our sample. We also label Kepler-10 b, another possible thick disk planet.}
\label{fig:figure1}
\end{figure*}

\begin{deluxetable*}{llcrlrr}
\setlength{\tabcolsep}{1em}
\tablewidth{0.9\textwidth}
\tabletypesize{\footnotesize}
\tablewidth{0pt}
\tablecaption{Host Star Properties}
\tablecolumns{7}
\tablehead{
\colhead{Name} &
\colhead{$T_{\textrm{eff}}$} &
\colhead{log$g$} &
\colhead{$M_{*}$} &
\colhead{$R_{*}$} &
\colhead{[Fe/H]} &
\colhead{[Mg/Fe]} \\[-0.2cm]
\colhead{} &
\colhead{K} &
\colhead{dex} &
\colhead{$M_{\odot}$} &
\colhead{$R_{\odot}$} &
\colhead{dex} &
\colhead{dex} 
}
\startdata
EPIC 220674823 & 5448 $\pm$ 32 & 4.2 $\pm$ 0.03 & 0.90 $\pm$ 0.01 & 0.97 $\pm$ 0.02 & 0.07 $\pm$ 0.01 & $-$0.01 $\pm$ 0.02 \\
TOI-561 & 5318 $\pm$ 12 & 4.5 $\pm$ 0.03 & 0.73 $\pm$ 0.01 & 0.83 $\pm$ 0.01 & $-$0.36 $\pm$ 0.01 & 0.22 $\pm$ 0.02 \\
TOI-1798 & 5093 $\pm$ 14 & 4.5 $\pm$ 0.03 & 0.83 $\pm$ 0.01 & 0.78 $\pm$ 0.01 & 0.00 $\pm$ 0.01 & 0.02 $\pm$ 0.02 \\
K2-38 & 5591 $\pm$ 19 & 4.1 $\pm$ 0.03 & 1.01 $\pm$ 0.01 & 1.15 $\pm$ 0.02 & 0.24 $\pm$ 0.01 & 0.01 $\pm$ 0.02 \\
Kepler-10 & 5699 $\pm$ 32 & 4.3 $\pm$ 0.04 & 0.91 $\pm$ 0.01 & 1.07 $\pm$ 0.02 & $-$0.11 $\pm$ 0.01 & 0.09 $\pm$ 0.02 \\
Kepler-21 & 6004 $\pm$ 12 & 3.8 $\pm$ 0.03 & 1.18 $\pm$ 0.02 & 1.95 $\pm$ 0.03 & $-$0.15 $\pm$ 0.01 & $-$0.05 $\pm$ 0.02 \\
Kepler-323 & 5880 $\pm$ 21 & 4.2 $\pm$ 0.03 & 0.95 $\pm$ 0.01 & 1.13 $\pm$ 0.02 & $-$0.19 $\pm$ 0.01 & 0.00 $\pm$ 0.02 \\
Kepler-93 & 5619 $\pm$ 7 & 4.3 $\pm$ 0.03 & 0.88 $\pm$ 0.00 & 0.93 $\pm$ 0.01 & $-$0.14 $\pm$ 0.01 & 0.02 $\pm$ 0.01 \\
K2-265 & 5342 $\pm$ 21 & 4.2 $\pm$ 0.03 & 0.85 $\pm$ 0.01 & 0.93 $\pm$ 0.01 & 0.00 $\pm$ 0.01 & 0.02 $\pm$ 0.02 \\
K2-291 & 5477 $\pm$ 7 & 4.3 $\pm$ 0.03 & 0.94 $\pm$ 0.01 & 0.90 $\pm$ 0.01 & 0.11 $\pm$ 0.01 & $-$0.05 $\pm$ 0.01 \\
K2-141 & 4693 $\pm$ 42 & 4.6 $\pm$ 0.04 & 0.74 $\pm$ 0.01 & 0.69 $\pm$ 0.01 & 0.04 $\pm$ 0.03 & 0.03 $\pm$ 0.05 \\
TOI-1444 & 5378 $\pm$ 36 & 4.5 $\pm$ 0.03 & 0.89 $\pm$ 0.02 & 0.91 $\pm$ 0.02 & 0.07 $\pm$ 0.03 & 0.05 $\pm$ 0.04 \\
Kepler-100 & 5766 $\pm$ 13 & 4.0 $\pm$ 0.03 & 1.08 $\pm$ 0.02 & 1.55 $\pm$ 0.03 & 0.07 $\pm$ 0.01 & 0.02 $\pm$ 0.02 \\
Kepler-407 & 5449 $\pm$ 28 & 4.1 $\pm$ 0.03 & 1.00 $\pm$ 0.01 & 1.02 $\pm$ 0.03 & 0.37 $\pm$ 0.01 & 0.02 $\pm$ 0.02 \\
K2-216 & 4541 $\pm$ 11 & 4.6 $\pm$ 0.03 & 0.68 $\pm$ 0.01 & 0.69 $\pm$ 0.02 & $-$0.05 $\pm$ 0.01 & 0.05 $\pm$ 0.02 \\
TOI-286 & 5022 $\pm$ 11 & 4.4 $\pm$ 0.03 & 0.79 $\pm$ 0.01 & 0.79 $\pm$ 0.01 & $-$0.03 $\pm$ 0.01 & 0.02 $\pm$ 0.02 \\
CoRoT-7 & 5241 $\pm$ 36 & 4.4 $\pm$ 0.03 & 0.87 $\pm$ 0.02 & 0.83 $\pm$ 0.03 & 0.07 $\pm$ 0.01 & 0.01 $\pm$0.02 \\
TOI-1347 & 5415 $\pm$ 38 & 4.5 $\pm$ 0.04 & 0.89 $\pm$ 0.01 & 0.84 $\pm$ 0.02 & 0.00 $\pm$ 0.01 & $-$0.03 $\pm$ 0.02 \\
TOI-1685 & 3449 $\pm$ 6 & -- & 0.41 $\pm$ 0.01 & 0.43 $\pm$ 0.003 & 0.05 $\pm$ 0.02 & 0.06 $\pm$ 0.02 \\
TOI-1450 & 3422 $\pm$ 6 & -- & 0.41 $\pm$ 0.01 & 0.43 $\pm$ 0.003 & 0.15 $\pm$ 0.02 & 0.03 $\pm$ 0.02 \\
TOI-1235 & 4083 $\pm$ 33 & -- & 0.62 $\pm$ 0.02 & 0.64 $\pm$ 0.01 & 0.00 $\pm$ 0.02 & 0.07 $\pm$ 0.03 \\
Wolf 327 & 3603 $\pm$ 46 & -- & 0.44 $\pm$ 0.01 & 0.46 $\pm$ 0.01 & 0.00 $\pm$ 0.02 & 0.04 $\pm$ 0.03 
\enddata
\tablecomments{This table lists the properties of the FGK and M dwarf stars that host our rocky planet sample. The $T_{\textrm{eff}}$ and log$g$ values come from ASPCAP, but ASPCAP log$g$ is unreliable for M dwarfs and so is unreported for the M dwarf hosts. The host star masses and radii were derived with \texttt{isoclassify}. The [Fe/H] and [Mg/Fe] abundances come from ASPCAP and the \citet{behmard2025} M dwarf catalog, for FGK and M dwarfs, respectively.}
\end{deluxetable*} \label{tab:table1}

\begin{deluxetable*}{lclrrrll}
\setlength{\tabcolsep}{1em}
\tablewidth{0.99\textwidth}
\tabletypesize{\footnotesize}
\tablewidth{0pt}
\tablecaption{Rocky Planet Properties}
\tablecolumns{8}
\tablehead{
\colhead{Name} &
\colhead{$R_{P}/R_{*}$} &
\colhead{K} &
\colhead{Radius} &
\colhead{Mass} &
\colhead{$\rho$} &
\colhead{P} &
\colhead{$e$} \\[-0.2cm]
\colhead{} &
\colhead{} &
\colhead{m s$^{-1}$} &
\colhead{$R_{\oplus}$} &
\colhead{$M_{\oplus}$} &
\colhead{$\rho_{\oplus}$} &
\colhead{days} &
\colhead{}
}
\startdata
EPIC 220674823 b & 0.01553 $\pm$ 0.00029$^{G}$ & 6.36 $\pm$ 0.57$^{G}$ & 1.65 $\pm$ 0.04 & 7.75 $\pm$ 0.71 & 9.54 $\pm$ 1.05 & 0.57 $\pm$ 0.0000055 & 0 \\
TOI-561 b & 0.01519 $\pm$ 0.00028$^{GP}$ & 1.95 $\pm$ 0.21$^{GP}$ & 1.38 $\pm$ 0.03 & 1.89 $\pm$ 0.21 & 3.97 $\pm$ 0.51 & 0.45 $\pm$ 0.0000003 & 0\\
TOI-1798.02 & 0.01622 $\pm$ 0.00076$^{AP}$ & 5.24 $\pm$ 0.66$^{AP}$ & 1.39 $\pm$ 0.07 & 5.51 $\pm$ 0.70 & 11.4 $\pm$ 2.18 & 0.44 $\pm$ 0.0000014 & 0\\
K2-38 b & 0.01330 $\pm$ 0.00076$^{TP}$ & 3.02 $\pm$ 0.43$^{B}$ & 1.66 $\pm$ 0.10 & 7.52 $\pm$ 1.06 & 9.03 $\pm$ 2.09 & 4.02 $\pm$ 0.0005 & 0.11\\
Kepler-10 b & 0.012684 $\pm$ 0.000041$^{D}$ & 2.58 $\pm$ 0.24$^{CB}$ & 1.48 $\pm$ 0.03 & 3.58 $\pm$ 0.34 & 6.12 $\pm$ 0.68 & 0.84 $\pm$ 0.0000003 & 0\\
Kepler-21 b & 0.007885 $\pm$ 0.00005$^{LM}$ & 2.70 $\pm$ 0.46$^{B}$ & 1.68 $\pm$ 0.03 & 6.68 $\pm$ 1.17 & 7.81 $\pm$ 1.42 & 2.79 $\pm$ 0.0000003 & 0\\
Kepler-323 c & 0.012849 $\pm$ 0.000261$^{M}$ & 2.78 $\pm$ 0.74$^{AT}$ & 1.59 $\pm$ 0.04 & 6.41 $\pm$ 1.67 & 8.79 $\pm$ 2.46 & 3.55 $\pm$ 0.000005 & 0\\
Kepler-93 b & 0.015858 $\pm$ 0.000809$^{M}$ & 1.89 $\pm$ 0.21$^{B}$ & 1.61 $\pm$ 0.09 & 4.55 $\pm$ 0.50 & 6.02 $\pm$ 1.25 & 4.73 $\pm$ 0.00000097 & 0\\
K2-265 b & 0.01695 $\pm$ 0.00095$^{T}$ & 3.87 $\pm$ 0.77$^{T}$ & 1.72 $\pm$ 0.10 & 7.17 $\pm$ 1.49 & 7.77 $\pm$ 2.21 & 2.37 $\pm$ 0.000058 & 0.16\\
K2-291 b & 0.01603 $\pm$ 0.00037$^{D}$ & 3.31 $\pm$ 0.56$^{D}$ & 1.57 $\pm$ 0.04 & 6.49 $\pm$ 1.10 & 9.21 $\pm$ 1.68 & 2.23 $\pm$ 0.000066 & 0\\
K2-141 b & 0.01993 $\pm$ 0.00052$^{D}$ & 6.10 $\pm$ 0.39$^{B}$ & 1.50 $\pm$ 0.05 & 5.11 $\pm$ 0.33 & 8.29 $\pm$ 0.94 & 0.28 $\pm$ 0.0000015 & 0\\
TOI-1444 b & 0.014277 $\pm$ 0.000404$^{AP}$ & 2.85 $\pm$ 0.37$^{CB}$ & 1.42 $\pm$ 0.05 & 3.24 $\pm$ 0.42 & 6.29 $\pm$ 1.08 & 0.47 $\pm$ 0.0000011 & 0\\
Kepler-100 b & 0.008062 $\pm$ 0.001336$^{M}$ & 1.25 $\pm$ 0.15$^{CB}$ & 1.36 $\pm$ 0.17 & 3.92 $\pm$ 0.45 & 8.60 $\pm$ 4.25 & 6.89 $\pm$ 0.0000067562 & 0\\
Kepler-407 b & 0.010404 $\pm$ 0.000284$^{M}$ & 1.41 $\pm$ 0.29$^{CB}$ & 1.15 $\pm$ 0.04 & 1.93 $\pm$ 0.53 & 6.95 $\pm$ 2.07 & 0.67 $\pm$ 0.0000005683 & 0\\
K2-216 b & 0.022 $\pm$ 0.0202$^{CP,AH}$ & 3.77 $\pm$ 1.00$^{AH}$ & 1.65 $\pm$ 0.17 & 5.94 $\pm$ 1.63 & 7.27 $\pm$ 3.32 & 2.17 $\pm$ 0.000056 & 0\\
TOI-286 b & 0.0167 $\pm$ 0.00117$^{MH}$ & 1.98 $\pm$ 0.33$^{MH}$ & 1.44 $\pm$ 0.10 & 4.39 $\pm$ 0.70 & 8.14 $\pm$ 2.46 & 4.51 $\pm$ 0.0000031 & 0\\
CoRoT-7 b & 0.0172 $\pm$ 0.0011$^{D}$ & 4.29 $\pm$ 0.46$^{AJ}$ & 1.55 $\pm$ 0.11 & 5.88 $\pm$ 0.65 & 8.71 $\pm$ 2.30 & 0.85 $\pm$ 0.000000587 & 0\\
TOI-1347 b & 0.02039 $\pm$ 0.00072$^{R}$ & 7.74 $\pm$ 0.80$^{R}$ & 1.86 $\pm$ 0.08 & 10.8 $\pm$ 1.10 & 9.15 $\pm$ 1.46 & 0.85 $\pm$ 0.00000061 & 0\\
TOI-1685 b & 0.02956 $\pm$ 0.00061$^{JB}$ & 3.76 $\pm$ 0.39$^{JB}$ & 1.49 $\pm$ 0.05 & 3.07 $\pm$ 0.33 & 5.16 $\pm$ 0.76 & 0.67 $\pm$ 0.00000042 & 0\\
TOI-1450 b & 0.0219 $\pm$ 0.0006$^{MB}$ & 1.05 $\pm$ 0.1$^{MB}$ & 1.12 $\pm$ 0.04 & 1.26 $\pm$ 0.12 & 4.93 $\pm$ 0.77 & 2.04 $\pm$ 0.000001 & 0\\
TOI-1235 b & 0.02508 $\pm$ 0.00085$^{PB}$ & 3.87 $\pm$ 0.37$^{LP}$ & 1.71 $\pm$ 0.07 & 6.58 $\pm$ 0.65 & 7.19 $\pm$ 1.14 & 3.44 $\pm$ 0.00009 & 0.049\\
Wolf 327 b & 0.0280 $\pm$ 0.0007$^{FM}$ & 3.54 $\pm$ 0.62$^{FM}$ & 1.29 $\pm$ 0.05 & 2.57 $\pm$ 0.45 & 6.65 $\pm$ 1.39 & 0.57 $\pm$ 0.0000003 & 0
\enddata
\tablecomments{This table lists the properties our rocky planet sample. We homogeneously recalculate planet radii, masses, and bulk densities from host star parameters and literature $R_{P}/R_{*}$ and RV semi-amplitude $K$ values. The references for these values are: G = \citet{guenther2024}, GP = \citet{piotto2024}, AP = \citet{polanski2024}, TP = \citet{toledo_padron2020}, B = \citet{bonomo2023}, D = \citet{dai2019}, CB = \citet{brinkman2025}, LM = \citet{lopez_morales2016}, M = \citet{morton2016}, AT = \citet{thomas2024}, T = \citet{thygesen2023}, JB = \citet{burt2024}, MB = \citet{brady2024}, PB = \citet{bluhm2020}, LP = \citet{luque2022}, FM = \citet{murgas2024}, CP = \citet{persson2018}, AH = \citet{howard2025}, MH = \citet{hobson2024}, AJ = \citet{john2022}, R = \citet{rubenzahl2024}}
\end{deluxetable*} \label{tab:table2}

\begin{figure*}[t]
    \centering
    \includegraphics[width=0.98\textwidth]{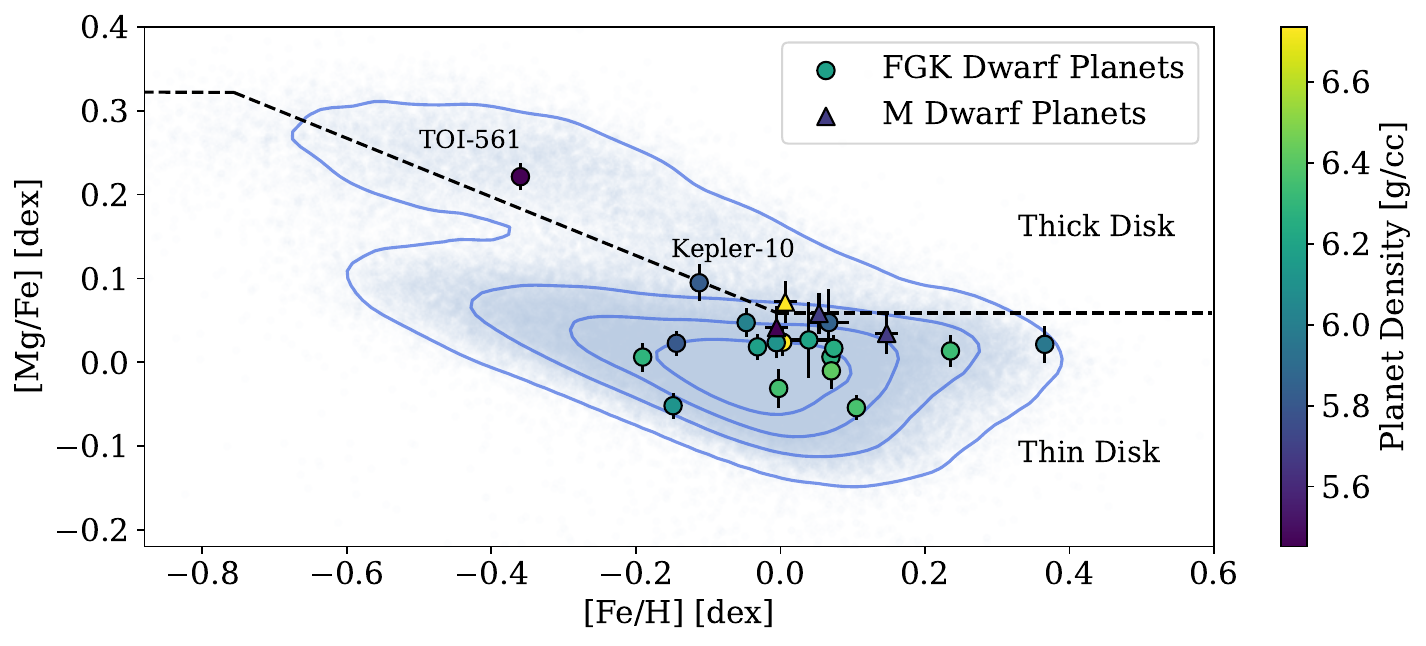} 
\caption{[Mg/Fe] vs. [Fe/H] for stars that host our rocky planet sample, as well as all main-sequence dwarfs in SDSS-V with reliable abundances. The SDSS-V dwarfs are plotted in the background as transparent blue dots, while the FGK and M dwarfs that host our rocky planet sample are marked as circles and triangles, respectively, and are colored by the densities of their rocky planets. The blue contours trace the background star 10\% density distribution levels, and the dashed black line delineates the thin-thick disk separation \citep{franchini2020}. TOI-561 and Kepler-10 are highlighted as high-confidence and possible thick disk stars, respectively.}
\label{fig:figure2}
\end{figure*}

\section{Sample Selection} \label{sec:sample}

\subsection{Rocky Planets}
We select small, rocky planets that have a high likelihood of lacking substantial volatile-rich or gaseous envelopes. This enables comparison between rocky planet interior properties, and host star abundances for refractory elements that compose the bulk of Earth's mantle and core (e.g., Fe, O, Si, Mg, etc.). To do this, we query the NASA Exoplanet Archive\footnote{\url{https://exoplanetarchive.ipac.caltech.edu/}, queried 7/25/2025} \citep{christiansen2025} and make an initial cut on planet radii ($R_{P}$ $\leq$ 1.8 $R_{\oplus}$). This leaves us with planets predominantly on the rocky side of the radius gap, which separates rocky planets from those with substantial gaseous envelopes, i.e., sub-Neptunes (e.g., \citealt{lopez2013,owen2013,fulton2017}). 

We then select rocky planets with robust radius and mass measurements by only retaining those with $R_{P}$/$R_{*}$ and RV semi-amplitude $K$ measurements that have $<$30\% fractional uncertainties. After these cuts, we are left with 81 planets. We ultimately only include planets with RV mass measurements, and exclude those with masses from transit timing variations (TTVs) because there may be discrepancies in the density distributions of RV vs. TTV planet samples (e.g., \citealt{mills2017,leleu2023,adibekyan2024}). We conduct a careful literature search to verify that the mass and radius measurements of each planet are the most up to date and/or inclusive of the most data, and were the result of a careful study often focused on each single planetary system. Finally, we apply cuts to only retain rocky planets with host stars that have well-measured elemental abundances. These cuts are detailed in Section \ref{sec:aspcap}. 

We check our rocky planets against the mass-radius models from \citet{zeng2019}, and find that all besides TOI-561 b fall within 1$\sigma$ of the rocky planet model bounds. TOI-561 b has been carefully characterized \citep{weiss2021,brinkman2023,lacedelli22}, and recent 3-5 $\mu$m JWST/NIRSpec observations provide brightness temperature constraints that can be explained by cooling from a volatile-rich atmosphere \citep{teske2025}. Thus, it is unclear if TOI-561 b can be classified as a rocky planet, but it is included in other rocky planet composition studies \citep{brinkman2024,brinkman2025}. It is also the only high-confidence thick disk planet in our sample, making it a valuable test case for investigating small planet compositions across different host star regimes. For these reasons, we keep TOI-561 b in our sample, and discuss it further in Section \ref{sec:discussion}.

\subsection{Host Star Abundances} \label{sec:aspcap}
We cross-match our rocky planet sample with Milky Way Mapper (MWM) data from the current phase of the Sloan Digital Sky Survey (SDSS-V) \citep{kollmeier2017}. We use SDSS-V/MWM and its associated data products to obtain elemental abundances for rocky planet host stars. The abundances are measured from high-resolution ($R$ = 22,500) stellar spectra in the $H$-band (1.51–1.7 $\mu$m), taken by the Apache Point Observatory Galactic Evolution Experiment (APOGEE) spectrographs \citep{majewski2017,wilson2019,almeida2023}. The spectra are processed with the APOGEE Stellar Parameter and Chemical Abundances Pipeline (ASPCAP) \citep{garcia2016,dr19}. ASPCAP makes use of MARCS model atmospheres \citep{gustafsson2008} and custom APOGEE line lists from \citet{shetrone2015} and \citet{smith2021} to generate synthetic spectra. It then uses the FERRE code \citep{allende_prieto2006} to interpolate within the synthetic spectral library for determining the best fit to an observed spectrum. ASPCAP provides abundances for a wide set of elements at typical precisions of $<$0.1 dex. We are interested in abundances for common elements in bulk Earth composition (e.g., \citealt{mcdonough2003}). Considering which of these elements are reliable from ASPCAP for dwarfs, we select X = Fe, O, Si, Mg, Ni, Ca, Al, and C. We impose the following cuts to ensure that the abundances are high quality:
  
\vspace{5mm}
\noindent 1. $T_{\textrm{eff}}$ \hspace{1mm}= 4500$-$6500 K
\vspace{0.8mm}
\newline \noindent 2. SNR $>$ 50
\vspace{0.8mm}
\newline \noindent 3. [X/H]$_{\textrm{error}}$ $<$ 0.1 dex
\vspace{0.8mm}
\newline \noindent 4. Flag \texttt{x\_h\_flag} not set
\vspace{5mm}

ASPCAP does not provide reliable abundance measurements for main-sequence stars with $<$4500 K temperatures, so we do not apply the above ASPCAP parameter cuts to select M dwarf planet hosts. Instead, we directly cross-match to the \citet{behmard2025} catalog of M dwarf abundances, which already includes $T_{\textrm{eff}}$ and SNR cuts. The SNR cut is the same as for the FGK dwarfs (SNR $>$ 50), and the $T_{\textrm{eff}}$ cut is 3000 K $<$ $T_{\textrm{eff}}$ $<$ 4000 K. The \citet{behmard2025} catalog provides abundances (Fe, Mg, Al, Si, C, N, O, Ca, Ti, Cr, and Ni) for $\sim$17,000 M dwarfs, and is based on a data-driven model and training set composed of SDSS-V/MWM FGK-M dwarf binaries. By assuming that binary companions share a parent molecular cloud and are thus born approximately chemically homogeneous (e.g., \citealt{de_silva2007,de_silva2009,bland_hawthorn2010}), we were able to tag the M dwarfs with the ASPCAP abundances of their FGK dwarf companions. The training set ultimately consisted of M dwarf spectra and abundance ``labels" from the FGK dwarfs. For more details on the data-driven model and resulting M dwarf abundance catalog, see \citet{behmard2025}. 

After applying the ASPCAP parameter cuts to our sample of rocky planets with FGK dwarf hosts, it decreases from 81 to 17 planets. Cross-matching our sample of rocky planets around M dwarf hosts with the \citet{behmard2025} catalog yields four planets. We provide the host star and planet information of our sample in Tables \ref{tab:table1} and \ref{tab:table2}, respectively. To get a sense of our host star abundances in terms of Galactic chemical evolution, we show where they fall within [Mg/Fe] vs. [Fe/H] space in Figure \ref{fig:figure2}.

Though TOI-1347 b does not pass the ASPCAP abundance cuts for all considered elements (fails for Si, O, and Ni), we retroactively add it to our sample because it is well-characterized and has reliable host star abundances for Fe and Mg \citep{rubenzahl2024}. These elements end up being the most important for our investigation. This brings our rocky planet sample count to 18 planets around FGK dwarfs, and four planets around M dwarfs. Our final sample of 22 rocky planets is illustrated in Figure \ref{fig:figure1}.



\begin{figure*}[t]
    \centering
    \includegraphics[width=0.98\textwidth]{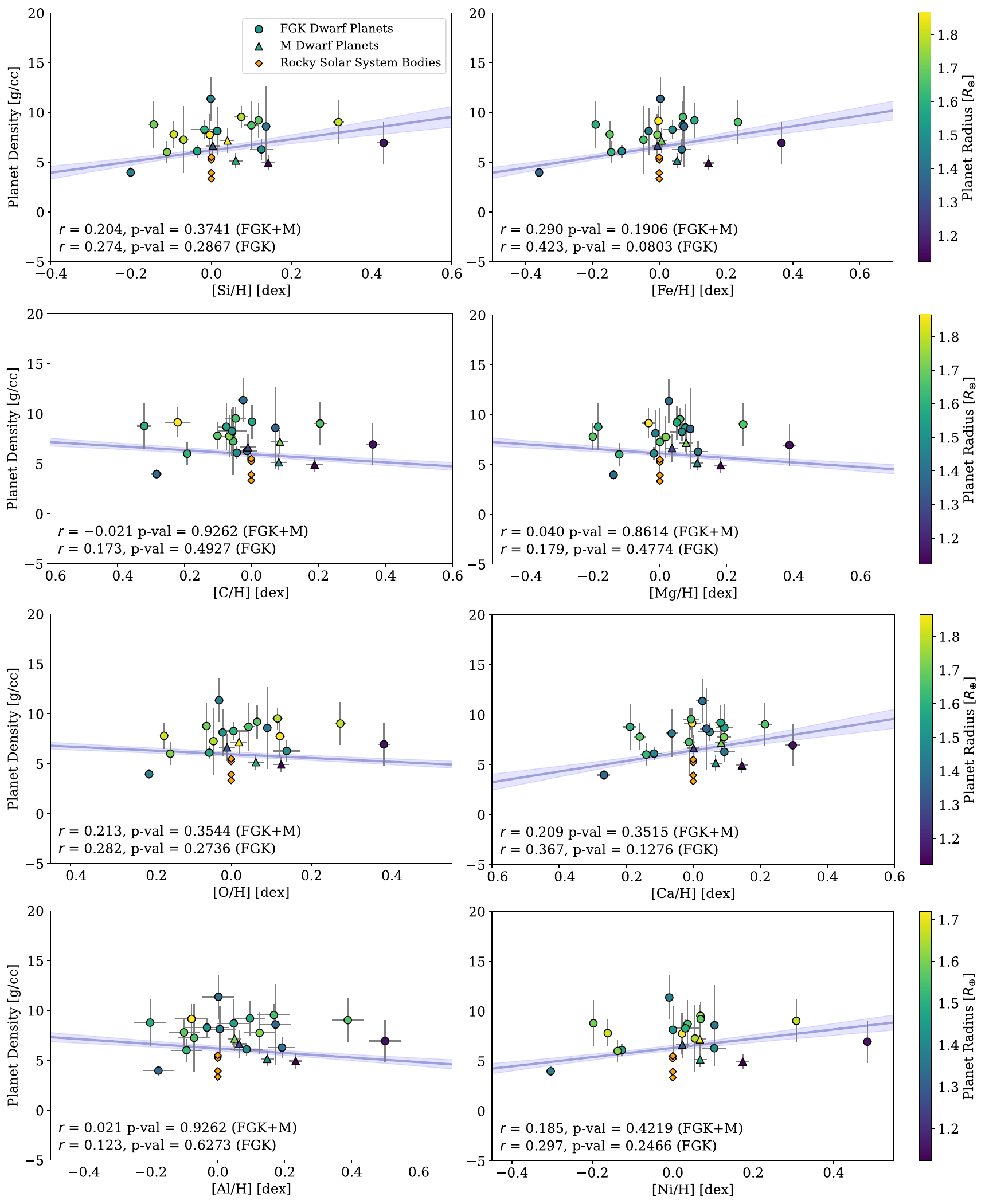} 
    \caption{Planet bulk density vs. all rocky elements considered in our analysis in [X/H] form. Rocky solar system bodies (Moon, Mercury, Venus, Earth, and Mars) are plotted for comparison as orange diamonds. Rocky planets hosted by FGK and M dwarfs are denoted as circles and triangles, respectively, and are colored by planet radii values. The best-fit linear trends from our MCMC fitting routine are plotted as blue lines along with their 1$\sigma$ confidence intervals. We provide the Pearson correlation statistics in the bottom left corners of each panel considering just the FGK-hosted rocky planets, and the entire FGKM sample. TOI-1347 b is not present in the Si, O, and Ni panels because it has unreliable host star abundances for these elements.}
    \label{fig:figure3}
    \end{figure*}

\begin{figure*}[t]
    \centering
    \includegraphics[width=0.988\textwidth]{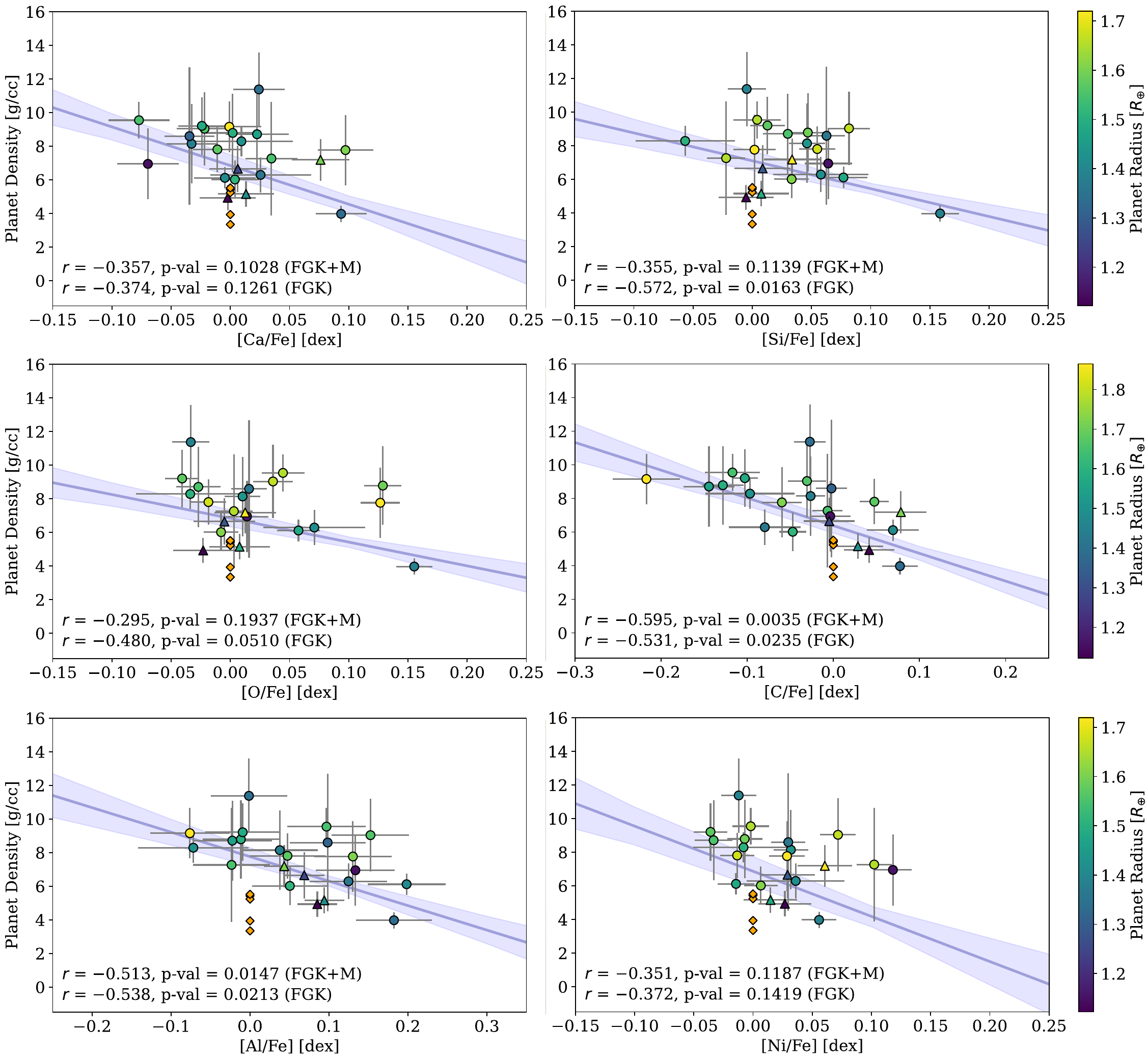} 
\caption{Planet bulk density vs. all rocky elements considered in our analysis in [X/Fe] form. As in Figure \ref{fig:figure3}, rocky solar system bodies (Moon, Mercury, Venus, Earth, and Mars) are plotted for comparison as orange diamonds. Rocky planets hosted by FGK and M dwarfs are denoted as circles and triangles, respectively, and are colored by planet radii values. The best-fit linear trends from our MCMC fitting routine are plotted as blue lines along with their 1$\sigma$ confidence intervals. We provide the Pearson correlation statistics in the bottom left corners of each panel considering just the FGK-hosted rocky planets, and the entire FGKM sample. TOI-1347 b is not present in the Si, O, and Ni panels because it has unreliable host star abundances for these elements.}
\label{fig:figure4}
\end{figure*}

\subsection{Updated Planet and Host Star Properties}

We homogeneously re-derive planet masses and radii for our rocky planet sample using their reported values for $R_{P}/R_{*}$ and RV semi-amplitudes $K$. This requires homogeneous host star masses and radii, which we measure with the \texttt{isoclassify} package \citep{huber2017}. Using MIST isochrone grids, \texttt{isoclassify} probabilistically infers stellar parameters given photometric, spectroscopic, or other input information \citep{choi2016}. \texttt{isoclassify} also includes empirical relations for measuring M dwarf masses and radii as a function of 2MASS $K_{S}$ band magnitudes and [Fe/H] \citep{mann2015,mann2019}, extending its use to the M dwarf regime.

We first run \texttt{isoclassify} in ``direct mode" to calculate host star luminosity. For the FGK dwarfs, \texttt{isoclassify} input consists of $T_{\textrm{eff}}$, log$g$, and [Fe/H] from ASPCAP; parallaxes from \emph{Gaia} DR3; 2MASS $JHK$-band magnitudes; a 3D dust map; and bolometric corrections. For the M dwarfs, we provide the same input parameters except log$g$, because ASPCAP does not provide accurate M dwarf log$g$ values (e.g., \citealt{souto2022}). We also draw M dwarf [Fe/H] values from the \citet{behmard2025} M dwarf catalog rather than ASPCAP. We then use the derived luminosities, $T_{\textrm{eff}}$, and log$g$ from the ``direct mode" runs as input for subsequent \texttt{isoclassify} runs in ``grid mode" to infer stellar masses and radii. We apply this approach to all rocky planet FGK and M dwarf hosts in our sample, and report their luminosities, masses, and radii in Table \ref{tab:table1}. 

While including the \citet{mann2015,mann2019} relations extends \texttt{isoclassify} to the M dwarf regime, that does not necessarily mean \texttt{isoclassify} will provide more accurate M dwarf parameters compared to using the empirical relations directly (discussion with D. Huber). However, we opt to use \texttt{isoclassify} for our M dwarfs because we use it for our FGK dwarfs, and we want self-consistent properties across our entire host star sample. We check the \texttt{isoclassify} M dwarf masses and radii against values from using the \citet{mann2015,mann2019} relations directly, and find that they agree in mass to within $<$15\%, and in radius to within $<$9\%. Looking ahead, we note that using M dwarf mass and radius values from either \texttt{isoclassify} or the \citet{mann2015,mann2019} relations directly does not change our main result (a correlation between host star [Mg/Fe] and planet density with $p$ = 0.0005 significance).

Using the newly derived host star masses, we re-derive planet masses using literature values for RV semi-amplitude $K$, orbital period, inclination, and eccentricity with the relation: 

\begin{equation}
    K = \frac{28.4329 \hspace{0.5mm} \textrm{m} \hspace{0.5mm} \textrm{s}^{-1}}{\sqrt{1-e^{2}}} \frac{M_{P}\hspace{0.2mm}\textrm{sin}\hspace{0.2mm}i}{M_{Jup}} \left(\frac{M_{P} + M_{*}}{M_{\odot}}\right)^{-2/3} \left(\frac{P}{1 \hspace{0.5mm} \textrm{yr}}\right)^{-1/3}.
\end{equation}

We re-derive planet radii with our host star radii and literature values for $R_{P}/R_{*}$. Our updated planet masses and radii are provided in Table \ref{tab:table2}.

\begin{figure*}[t]
    \centering
    \includegraphics[width=0.94\textwidth]{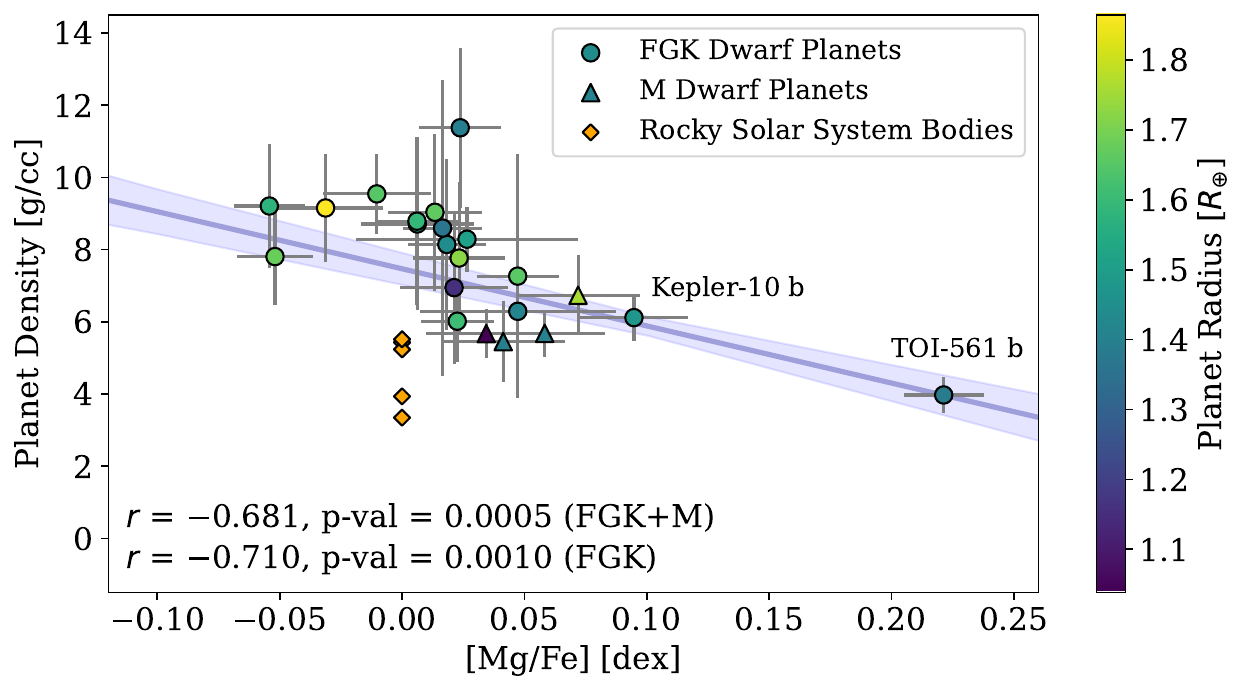} 
\caption{Rocky planet densities vs. host star [Mg/Fe] for planets hosted by both FGK (circles) and M dwarfs (triangles), colored by planet radii values. Rocky solar system bodies (Moon, Mercury, Venus, Earth, and Mars) are plotted for comparison as orange diamonds. Our best fit linear model from our MCMC fitting routine is plotted as a blue line along with the associated 1$\sigma$ confidence intervals. Pearson correlation statistics are provided in the bottom left corner considering just the FGK-hosted rocky planets, and the entire FGKM sample. We label the high-confidence and possible thick disk planets TOI-561 b and Kepler-10 b, respectively.}
\label{fig:figure5}
\end{figure*}

\section{Planet-Star Correlations} \label{sec:correlation}

We use our sample to investigate if rocky planet properties correlate with host star abundances of major rock-forming elements common in bulk Earth composition (i.e., Fe, O, Si, Mg, Ni, Ca, Al, and C) \citep{mcdonough2003}. We focus on planet properties close to observables (i.e., mass, radius, and bulk density) as opposed to secondary observables that depend on planet interior modeling, e.g., iron mass fraction or core mass fraction-like quantities (e.g., \citealt{schulze2021,adibekyan2021,brinkman2024}). This removes interior modeling-based uncertainties from our analysis, and frees us from assuming strictly Earth-like mineralogical compositions that such models are based on (e.g., \citealt{valencia2006,valencia2007,dorn2017}). 

We examine potential correlations between planet mass, radius, and bulk density vs. host star abundances in both [X/H] and [X/Fe] form. Because all abundances are correlated through nucleosynthetic pathways, the [X/Fe] form separates out the common component and enables us to more easily identify trends unique to each particular element. The [X/Fe] form also contains information on Galactic location for $\alpha$-elements X (e.g., Mg, Si, Ca), because [$\alpha$/Fe] systematically varies between different Galactic structural components such as the thin disk, thick disk, and halo. We find no correlations whatsoever between planet mass or radius vs. [X/H] or [X/Fe] host star abundances of any rock-forming element considered in our analysis. However, we do see potential trends between planet bulk density and host star abundances, particularly in [X/Fe] form. 

To model these trends, we carry out a linear regression fitting routine with Markov Chain Monte Carlo (MCMC) using \texttt{NumPyro} \citep{phan2019,bingham2019}, a probabilistic programming library based on \texttt{JAX} \citep{bradbury2018}. \texttt{JAX} provides gradients through autodifferentiation, which allows for efficient gradient-based MCMC algorithms such as Hamiltonian Monte Carlo and its variant No-U-Turn Sampling (NUTS) \citep{NUTS}. To construct the likelihood, we re-parameterize the model in terms of $\theta$, the angle between the fitted line and the horizontal axis. Then, instead of fitting for the slope $m$ and intercept $b$, we fit for $\theta = \textrm{arctan}\hspace{0.5mm}m$ and the perpendicular intercept $b_{\perp} = b\hspace{0.5mm}cos\hspace{0.5mm}\theta$. This re-parameterization enables us to use a uniform prior on $\theta$. This would not be possible with $m$ because uniform priors on $m$ favor steeper slopes, and could thus bias our results (e.g., \citealt{jaynes1991straight}). In contrast, uniform priors are minimally informative for $\theta$. 

We sample the posterior space using NUTS with two chains of 10,000 draws each, preceded by 10,000 warm-up steps. To check for convergence, we calculate the Gelman-Rubin $\hat{r}$ statistic that quantifies the variation between the two chains and within each single chain for each parameter \citep{gelman_rubin1992}. We check for convergence on slope and intercept, defined as $\hat{r}$ within 0.1 of 1. We show the fitted linear trends from this analysis for planet density vs. [X/H] and [X/Fe] in Figures \ref{fig:figure3} and \ref{fig:figure4}, respectively. 

We find that the trends between rocky planet density and host star abundances in [X/H] form are not significant and relatively flat, with slopes ranging from $m$ = $-$2.03 $\pm$ 0.55 to $m$ = 5.21 $\pm$ 1.20. The trends between planet density and host star abundances in [X/Fe] form are all negative and steeper, with slopes ranging from $m$ = ($-$26.9 $\pm$ 8.28 to $m$ = $-$14.1 $\pm$ 4.28. This inverse relationship makes sense if we assume that rocky planets have approximately Earth-like compositions, with iron-dominated cores and silicate mantles; if primordial planet-forming material is iron-depleted relative to other rocky elements, rocky planets form with smaller cores and thicker mantle layers, leading to lower bulk densities.

To quantify the trends between planet density and host star abundances, we compute Pearson correlation coefficients. We do not find any statistically significant correlations between density vs. host star abundances in [X/H] form, agreeing with the relatively flat slopes from MCMC linear fits. However, we do find significant correlations between planet density and abundances in [X/Fe] form. If we consider only the FGK-hosted rocky planets in our sample, we find that [Mg/Fe] and [Si/Fe] exhibit statistically significant trends with planet density ($p$ = 0.001 and 0.0163, respectively). When we add the four M dwarf-hosted planets in our sample, the [Mg/Fe] correlation significance increases to $p$ = 0.0005, but the correlation with [Si/Fe] decreases to $p$ = 0.1139. Conversely, correlations between bulk density and [C/Fe] increase from marginally significant ($p$ = 0.0235) to significant ($p$ = 0.0035) when considering rocky planets around just FGK hosts vs. the entire FGKM sample. Correlations between [Al/Fe] and planet density are marginally significant, and hardly change if we consider just FGK hosts, or all FGKM hosts ($p$ = 0.0213 and 0.0147, respectively).

Of all host star abundances considered, only [Mg/Fe] exhibits a correlation with rocky planet density that remains strongly significant whether we consider just FGK or all FGKM host stars (Figure \ref{fig:figure5}). Mg is a well-known $\alpha$-element purely produced in core collapse supernovae. Ratios of $\alpha$-to-Fe abundances delineate the Galactic thin and thick disks, which are characterized by low and high [$\alpha$/Fe], respectively. Thus, it is unsurprising that the [Mg/Fe] vs. planet density trend is anchored by the single high-confidence thick disk planet in our sample (TOI-561 b), whose thick disk status has been confirmed through host star abundances (e.g., [Mg/Fe] = 0.22 $\pm$ 0.02 dex) and kinematics (e.g., \citealt{weiss2021,brinkman2023}), and which clearly lies in the thick disk region of [Mg/Fe] vs. [Fe/H] space (Figure \ref{fig:figure2}). When we remove this planet, the correlation remains significant, but less so ($p$ = 0.0005 to 0.007). Our sample also includes Kepler-10 b, whose host star may also reside in the thick disk based on its abundances ([Mg/Fe] = 0.09 $\pm$ 0.02 dex) and kinematics \citep{batalha2011,brinkman2024}, but whose thick disk membership is much more unclear given its location in [Mg/Fe] vs. [Fe/H] space (Figure \ref{fig:figure2}). When we remove Kepler-10 b as well, the correlation is still significant at $p$ = 0.015. Thus, the trend between rocky planet density and host star [Mg/Fe] appears to hold in the absence of thick disk planets, but is greatly strengthened when we include them. Our small sample of thick disk planets limits our ability to draw firm conclusions about the properties of planets around thick-disk and/or $\alpha$-rich stars, but the cases we do see support the hypothesis that they have lower densities.

\begin{figure}[t]
    \centering
    \hspace{-4mm}
    \includegraphics[width=0.49\textwidth]{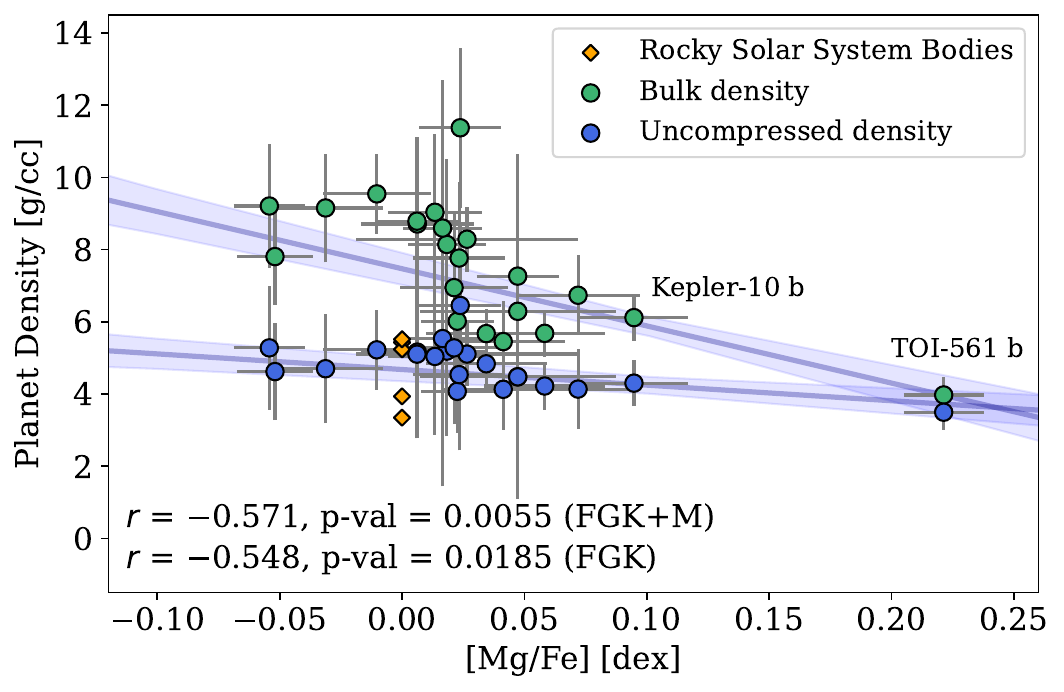} 
\caption{Uncompressed (blue) and bulk (green) rocky planet densities vs. host star [Mg/Fe] for planets hosted by both FGK (circles) and M dwarfs (triangles), colored by planet radii values. Rocky solar system bodies (Moon, Mercury, Venus, Earth, and Mars) are plotted for comparison as orange diamonds. Our best fit linear models from our MCMC fitting routine are plotted as blue lines along with their associated 1$\sigma$ confidence intervals. Pearson correlation statistics for the uncompressed density case are provided in the bottom left corner considering just the FGK-hosted rocky planets, and the entire FGKM sample. We label the high-confidence and possible thick disk planets TOI-561 b and Kepler-10 b, respectively.}
\label{fig:figure6}
\end{figure}

\subsection{Uncompressed Planet Densities}
We also compare uncompressed rocky planet densities to host star [Mg/Fe]. The uncompressed density of a planet is the density it would have if gravity did not compress its interior, which leads to higher bulk densities for larger planets. Thus, uncompressed densities more directly probe planet compositions. For example, the Earth and Mercury have similar bulk densities (5.5 and 5.4 g/cc, respectively), while their uncompressed densities belay their compositional differences (4.7 and 6.3 g/cc, respectively). Unfortunately, we cannot calculate the true uncompressed densities of our rocky planets because we require knowledge of what rock-building minerals they are made of. 

We approximate uncompressed densities by assuming our rocky planets have Earth-like interior properties. We do this with the \texttt{SuperEarth} package \citep{valencia2006, plotnykov2020}, which solves equations of state for iron and rock-building minerals to match the planet mass and radius values provided. \texttt{SuperEarth} assumes an Earth-like interior structure of an iron core and rocky mantle, composed of Earth-specific minerals (e.g., olivene, pyroxene, wadsleyite, ringwoodite, etc.). \texttt{SuperEarth} allows us to input the amount of iron in the mantle and silica in the core, which we set to the default Earth-like values of 0.1 iron mole fractions and no silica inclusions, respectively \citep{1995ChGeo.120..223M}. \texttt{SuperEarth} tunes the mantle vs. core contributions to match the planet mass and radius, then uncompresses the core and mantle to nominal reference density values of $\rho_{c}$ = 8.2 g/cc and $\rho_{c}$ = 4.0 g/cc \citep{plotnykov2020} to compute the overall uncompressed planet density. 

Assuming that rocky exoplanets are composed of the same minerals found on Earth is a big assumption, so \texttt{SuperEarth} can only provide an approximation of the true uncompressed densities. Still, we use it to investigate the trend between uncompressed planet density and host star [Mg/Fe]. We find it to be less strong compared to the trend with bulk density, but still significant ($p$ = 0.0055) (Figure \ref{fig:figure6}). The fact that this correlation remains strong even for imperfectly measured uncompressed densities indicates that it is quite robust.

\section{Discussion} \label{sec:discussion}

\subsection{Planet Formation in the Thin vs. Thick Disk}
In this study, we investigated links between rocky planet properties and the elemental abundances of their FGK and M dwarf host stars. We uncovered a significant correlation between host star [Mg/Fe] and rocky planet bulk density ($p$ = 0.0005). This may point to differences in rocky planet compositions depending on where planets form within the Galaxy, i.e., the thin vs. thick disk, which are characterized by lower and higher [$\alpha$/Fe] enrichment, respectively, for $\alpha$-elements like Mg. In accordance with this, the single high-confidence thick disk planet in our sample (TOI-561 b) strongly contributes to the observed correlation. The lower confidence thick disk planet in our sample (Kepler-10 b) supports the trend as well. 

If rocky planet densities vary strongly as a function of Galactic formation location, we might expect to find significant correlations with $\alpha$-elements beyond Mg, such as Si or Ca. We do see a tentative correlation between planet density and [Si/Fe] if we consider only the 18 planets with FGK hosts ($p$ = 0.0163), but this trend becomes insignificant with the addition of the four M dwarf-hosted planets in our sample ($p$ = 0.1139). We also do not find a statistically significant correlation between planet density and host star [Ca/Fe] (Figure \ref{fig:figure4}). However, we may only see a statistically significant correlation with Mg because it is, in a sense, the purest $\alpha$-element. Mg is purely created in core-collapse supernovae (CCSNe), as opposed to Si and Ca that have partial contributions from Type Ia supernovae (SNe Ia) (e.g., \citealt{million2011}). The division between CCSNe and SNe Ia nucleosynthetic channels characterizes the [$\alpha$/Fe] vs. [Fe/H] separation of the Galactic thin and thick disks. More specifically, CCSNe and SNe Ia are fiducial sources of Fe-peak and $\alpha$-elements, respectively, but SNe Ia are delayed by $\sim$0.04–10 Gyr after star formation events \citep{maza1976}. This delay in iron enrichment leads to a decrease in [$\alpha$/Fe] with increasing [Fe/H], and causes the younger Galactic thin disk to be relatively Fe-enhanced and $\alpha$-depleted compared to the older Galactic thick disk. Because Mg is a purer product of CCSNe, it is subsequently a purer $\alpha$-element for delineating the thin-thick disk boundary. Thus, the link between rocky planet density and host star [Mg/Fe] suggests that rocky planet compositions vary between the thin and thick disks, even in the absence of additional $\alpha$-element trends. 

To confirm the connection between rocky planet composition and Galactic formation location, we need more well-characterized thick disk planets. Our sample includes only one high-confidence thick disk planet (TOI-561 b), and one potential thick disk planet (Kepler-10 b). Both help anchor the correlation between planet density and host star [Mg/Fe], particularly TOI-561 b. Additional thick disk planets can corroborate this trend, but detecting them is challenging---thick disk planets are intrinsically more rare than their thin disk counterparts. \citet{zink2023} report a 50\% $\pm$ 8\% reduction in super-Earth occurrence within orbital periods of $\sim$40 days around thick disk stars. Because planet formation favors metal-rich environments, it may be subdued in the thick disk where stars are metal-poor ($\sim$\hspace{0.5mm}$-$0.5 dex, e.g., \citealt{reddy2006,kordopatis2011,hayden2015}). Current planet surveys (e.g., TESS) have begun targeting stars in the thick disk and halo, but detecting planet signatures around these stars remains difficult because they are so faint. Luckily, upcoming large scale surveys such as \emph{Gaia} DR4 and the PLAnetary Transits and Oscillations of stars (PLATO)
mission will enable us to measure ages, abundances, and other properties of many more planet host stars in thick disk and halo regions (e.g., \citealt{gaia2023,boettner2024}).

As we discover more planets in the thick disk, it will be important to characterize them further with follow-up observations (e.g., JWST). While TOI-561 b was assumed to be rocky in previous studies (e.g., \citealt{weiss2021,brinkman2024,brinkman2025}), recent 3-5 $\mu$m JWST/NIRSpec observations provide brightness temperature constraints that are inconsistent with a bare rock surface, and can instead be explained by water-rich atmospheric compositions \citep{teske2025}. An atmosphere on TOI-561 b is unlikely to be primordial given the planet's low mass and high irradiation levels. Instead, it would likely stem from outgassing of volatile-rich species within the molten surface and/or iron core. An atmosphere would undoubtedly contribute to TOI-561 b's low bulk density. However, it remains an interesting and relevant planet in our study because its atmosphere and interior composition may be connected to its location within the thick disk. \citet{teske2025} speculate that TOI-561 b's interior refractory abundances reflect the $\alpha$-element-rich, iron-poor thick disk environment. Refractory abundance ratios (Fe/Mg, Fe/Si) in planet interiors are expected to impact volatile retention vs. outgassing (e.g., \citealt{lichtenberg2021,schlichting2022,bower2022}), and recent molecular dynamics simulations indicate that water strongly partitions into iron over silicates, and preferentially remains within iron-dominated cores \citep{luo2024}. TOI-561 b's formation within the thick disk may have led to low iron content and a small core that facilitated outgassing to form a water-rich atmosphere. Thus, planet location in the thin vs. thick disks may influence secondary atmosphere formation, but we need more carefully characterized thick disk planets to further investigate this possibility.

\subsection{Rocky Planets Around M Dwarfs}

Investigating the connection between rocky planet properties and host star chemistry is particularly important for M dwarf planet hosts; M dwarfs are the most common stars, have the largest RV and transit signals for small planets, and are orbited by planets $\sim$3x more often than FGK dwarfs (e.g., \citealt{dressing2015b,mulders2015,gaidos2016,hsu2020}). M dwarfs are also interesting because they are often old, due to their low masses that translate to slow hydrogen fusion rates at their cores (e.g., \citealt{woolf2020}). Thus, M dwarf chemical compositions are fossil records that encode nucleosynthetic processes and interstellar medium enrichment from the early days of our Galaxy \citep{bochanski2010,woolf2012}. Connecting their chemistry to rocky planet properties could yield unprecedented insight into how Earth-like planet formation relates to Galactic chemical evolution.

This is the first demographics study of rocky planet host star chemistry to include M dwarfs. Because constraining M dwarf chemical compositions is nontrivial, we source our sample from an M dwarf abundance catalog inferred from a data-driven approach \citep{behmard2025}, which yields only four M dwarf-hosted rocky planets that meet our sample criteria. Still, this small sample exhibits interesting properties compared to the FGK-hosted planets. All four planets around M dwarfs have low densities, with an average bulk density of 5.9 g/cc (uncompressed 4.3 g/cc), as compared to the average bulk density of planets around FGK dwarfs of 8.3 g/cc (uncompressed 5.0 g/cc). Thus, the M dwarf-hosted planets help anchor the lower end of the planet density-host star [Mg/Fe] correlation. The significance of this correlation gets stronger with the addition of these four planets ($p$ = 0.001 to 0.0005), indicating that it persists across the entire FGKM host star range.

Interestingly, several studies report discoveries of rocky planets around M dwarfs with lower densities compared to the solar system terrestrial planets (e.g., \citealt{agol2021,luque2022,cherubim2023,piaulet2023,tyler2025}), though it remains unclear if M dwarf-hosted rocky planets are systematically less dense than their FGK-hosted counterparts. If true, it would agree with the fact that M dwarfs are often relatively old, with super-solar [$\alpha$/Fe] abundance ratios from formation during earlier periods when CCSNe enrichment dominated the Galactic chemical landscape. To understand this, we will need larger M dwarf samples. Ongoing and future large scale spectroscopic surveys (e.g., SDSS-V, \emph{Gaia} DR4) will be invaluable for this.

\subsection{Are Rocky Exoplanet Compositions Earth-like?}

If we assume that rocky planets have broadly Earth-like compositions (i.e., iron-dominated cores surrounded by silicate-rich mantles), the trend we uncover between planet density and host star [Mg/Fe] has a fairly straightforward interpretation. That is, rocky planets form with larger, dense iron-dominated cores in more Fe-rich regions of the Galaxy, like the thin disk, and are more silicate-mantle dominated if they form in the thick disk, where $\alpha$-elements are more abundant compared to Fe. The thin and thick disk formation scenarios then produce rocky planets with higher and lower bulk densities, respectively. The planet density-host star [Mg/Fe] trend is consistent with this, and thus supports the idea that rocky exoplanets have approximately Earth-like interior compositions throughout the Galaxy, but with different mantle vs. core contributions.

Of course, a larger rocky planet sample is needed to further test this idea. In this study, we define rocky, Earth-like planets as those with $R_{P}$ $\leq$ 1.8 $R_{\oplus}$, which is regarded as the approximate location of the radius valley. However, more stringent radius cuts (e.g., $R_{P}$ $\leq$ 1.5 $R_{\oplus}$) exclude the valley altogether, only retaining the pile of smaller planets more confidently regarded as completely rocky, devoid of gaseous envelopes (e.g., \citealt{dressing2015,fulton2017}). With a larger sample that includes more small planets, we can test the [Mg/Fe] vs. planet density correlation in stricter rocky, Earth-like planet regimes. For now, we test more stringent radius cuts of $\leq$1.5 and 1.6 $R_{\oplus}$ on our current sample of 22 planets. They cut our sample down to 14 and 10 planets, respectively, and yield less significant correlations between host star [Mg/Fe] and planet density ($p$ = 0.005 and 0.056, respectively). While this decreasing significance may point to more complicated relationships between host star chemistry and planet properties for the smallest, rockiest planets, it also likely stems from having too few planets to confirm the correlation. More $\leq$1.5 $R_{\oplus}$ planets are needed to investigate this further. 

A larger sample will also provide more planets spread across different Galactic regions, which can probe the weaker correlations we see between planet density and other host star abundances, namely [Si/Fe], [Al/Fe], and [C/Fe]. If verified, these trends may point to terrestrial planet compositions that deviate from Earth-like. For example, the Earth is significantly carbon-depleted compared to interstellar medium solids and the solar nebula, with a C/Si ratio more than three orders of magnitude below solar values (e.g., \citealt{allegre2001,bergin2015}). It is unclear what mechanism is behind the carbon-depletion (e.g., \citealt{binkert2023}), and how widespread it might be in rocky planets throughout the Galaxy. 

There are hints that rocky exoplanet compositions may deviate from strictly Earth-like. For example, \citet{agol2021} used transit-timing variations and a photodynamical model to precisely characterize the seven rocky planets in TRAPPIST-1, and found that they adhere to a slightly different mass-radius relation compared to the solar system terrestrial planets, with molar Fe/Mg = 0.75 as compared to the solar value of Fe/Mg = 0.83. \citet{agol2021} conclude that the TRAPPIST-1 planets have consistently lighter interiors compared to the solar system rocky planets, which could be due to lower iron content or volatile enrichment. This is consistent with age estimates of TRAPPIST-1
(7.6$-$8 Gyr, \citealt{burgasser2017,birky2021}) that place it in the transitional thin/thick disk zone with enhanced [$\alpha$/Fe] abundance ratios. Thus, the TRAPPIST-1 planets appear to follow our uncovered correlation between planet density and host star [Mg/Fe]. However, a more precise analysis of TRAPPIST-1 is needed to determine if the planets follow an Earth-like interior breakdown of iron-dominated core and silicate-dominated mantle, are composed of more exotic refractory species, or have significant non-Earth-like volatile enrichment. So far, observations of TRAPPIST-1 b, c, and d indicate that these planets lack atmospheres \citep{zieba2023,greene2023,piaulet2025}, ruling out the possibility of atmospheric volatile enrichment.

As we begin to place the rocky planet population in a Galactic context, it is vital to disentangle the effects of host star type, chemistry, age, and environment dynamics. Many of these factors overlap with one another; M dwarfs live longer than solar-like stars and are often older, age and stellar chemistry are linked, etc. For example, \citet{weeks2025} report a connection between rocky planet density and stellar age, but ultimately attribute this effect to Galactic chemical evolution. Their result likely echoes our own, where stellar age measurements are a noisier representation of [$\alpha$/Fe] abundances for $\alpha$-elements like Mg. Still, system age may have independent effects on planet formation. Thick disk stars predominantly formed during ``cosmic noon" at redshift $z$ $\sim$ 2---a period of peak star formation in Galactic history \citep{madau2014,schreiber2020}. This turbulent environment caused protoplanetary disks to experience UV radiation at levels $\sim$7 orders of magnitude higher than in the solar neighborhood, curtailing their life to just $\sim$0.2$-$0.5 Myr and drastically limiting the timescale over which planets could form \citep{hallat2025}. Thus, many different properties of planet host stars can provide causal explanations for trends we see in the planet population. Our task is to distinguish between them, so we can understand which dominate in planet formation pathways across different regions of the Galaxy.



\section*{Acknowledgements}
We thank Shreyas Vissapragada, Catherine Manea, Heather Knutson, Dan Huber, Romy Rodr\'{i}guez Mart\'{i}nez, Caroline Piaulet, Madison Brady, and Lehman Garrison, as well as the CCA Astro Data and Nearby Universe groups for supportive and productive conversations. We also thank the anonymous referee for a helpful report. 

Funding for the Sloan Digital Sky Survey V has been provided by the Alfred P. Sloan Foundation, the Heising-Simons Foundation, the National Science Foundation, and the Participating Institutions. SDSS acknowledges support and resources from the Center for High-Performance Computing at the University of Utah. SDSS telescopes are located at Apache Point Observatory, funded by the Astrophysical Research Consortium and operated by New Mexico State University, and at Las Campanas Observatory, operated by the Carnegie Institution for Science. The SDSS web site is \url{www.sdss.org}.

SDSS is managed by the Astrophysical Research Consortium for the Participating Institutions of the SDSS Collaboration, including the Carnegie Institution for Science, Chilean National Time Allocation Committee (CNTAC) ratified researchers, Caltech, the Gotham Participation Group, Harvard University, Heidelberg University, The Flatiron Institute, The Johns Hopkins University, L'Ecole polytechnique f\'{e}d\'{e}rale de Lausanne (EPFL), Leibniz-Institut f\"{u}r Astrophysik Potsdam (AIP), Max-Planck-Institut f\"{u}r Astronomie (MPIA Heidelberg), Max-Planck-Institut f\"{u}r Extraterrestrische Physik (MPE), Nanjing University, National Astronomical Observatories of China (NAOC), New Mexico State University, The Ohio State University, Pennsylvania State University, Smithsonian Astrophysical Observatory, Space Telescope Science Institute (STScI), the Stellar Astrophysics Participation Group, Universidad Nacional Aut\'{o}noma de M\'{e}xico, University of Arizona, University of Colorado Boulder, University of Illinois at Urbana-Champaign, University of Toronto, University of Utah, University of Virginia, Yale University, and Yunnan University.

This work made use of data from the European Space Agency (ESA) mission Gaia (\url{https://www.cosmos.esa.int/gaia}), processed by the Gaia Data Processing and Analysis Consortium (DPAC, \url{https://www.cosmos.esa.int/web/gaia/dpac/consortium}). Funding for the DPAC has been provided by national institutions, in particular the institutions participating in the Gaia Multilateral Agreement.
\software{\texttt{numpy} \citep{numpy}, \texttt{matplotlib} \citep{matplotlib}, \texttt{pandas} \citep{pandas}, \texttt{scipy} \citep{scipy}, \texttt{scikit-learn} \citep{scikit-learn}, \texttt{astropy} \citep{astropy:2013, astropy:2018}, \texttt{JAX} \citep{bradbury2018}, \texttt{numpyro} \citep{phan2019,bingham2019}}

\bibliography{mybib}{}
\bibliographystyle{aasjournal}

\appendix
\setcounter{figure}{0}                       
\renewcommand\thefigure{A.\arabic{figure}}

\end{document}